\documentclass[twocolumn,showpacs,superscriptaddress,aps,prl]{revtex4}

\renewcommand{\paragraph}[1]{\textbf{#1.}---}

\usepackage{graphicx} % Include figure files
\usepackage{bm}       % bold math

\newcommand{\figurewidth}{86mm}
\bibpunct[]{[}{]}{,}{n}{}{,} % make multiple entries work properly

\begin{document}

\title{Universality Class of Criticality in the Restricted Primitive Model
Electrolyte}

\author{Erik Luijten}
\altaffiliation[Now at ]{Dept.\ of Materials Science and Engineering,
University of Illinois, Urbana, IL 61801}
\email[]{luijten@uiuc.edu}

\author{Michael E. Fisher}
\thanks{Corresponding author: fax (301) 314-9404.}
\affiliation{Institute for Physical Science and Technology, University of
Maryland, College Park, Maryland 20742}

\author{Athanassios Z. Panagiotopoulos}
\affiliation{Department of Chemical Engineering, Princeton University,
Princeton, New Jersey 08540}

\date{\today}

\begin{abstract}
  The 1:1 equisized hard-sphere electrolyte or \emph{restricted primitive
  model} has been simulated via grand-canonical fine-discretization Monte
  Carlo. Newly devised unbiased finite-size extrapolation methods using
  temperature-density, ($T, \rho$), loci of inflections, $Q \equiv \langle
  m^2\rangle^2/\langle m^4\rangle$ maxima, canonical and $C_V$ criticality,
  yield estimates of ($T_c$, $\rho_c$) to $\pm$(0.04, 3)\%.  Extrapolated
  exponents and $Q$-ratio are ($\gamma$, $\nu$, $Q_c$) = [1.24(3), 0.63(3);
  0.624(2)] which support Ising ($n \! = \! 1$) behavior with ($1.23_9$,
  $0.630_3$; $0.623_6$), but exclude classical, XY ($n \! = \!  2$), SAW ($n \!
  = \! 0$), and $n \!= \! 1$ criticality with potentials
  $\varphi(r)>\Phi/r^{4.9}$ when $r \to \infty$.
\end{abstract}


\maketitle

Since the experiments of Singh and Pitzer in 1988~\cite{fisher-review,weing},
an outstanding experimental and theoretical question has been: What is the
universality class of Coulombic criticality? Early experimental data for
electrolytes exhibiting phase separation driven by long-range ionic forces
suggested classical or van der Waals (vdW) critical behavior, with exponents
$\beta=\frac{1}{2}$, $\gamma=1$, $\nu=\frac{1}{2}$,
etc.~\cite{fisher-review,weing}: But the general theoretical consensus has been
that asymptotic Ising-type criticality, with $\beta \simeq 0.326$, $\gamma
\simeq 1.239$, $\nu \simeq 0.630$, etc., should be
expected~\cite{fisher-review,stell94,fisher-lee96}. Naively, one may argue that
the exponential Debye screening of the direct ionic forces results in
\emph{effective short-range} attractions that can cause separation into two
neutral phases: ion-rich and
ion-poor~\cite{fisher-review,weing,stell94,fisher-lee96,fisher-levin93}; the
order parameter, namely, the ion density or concentration difference, is a
scalar; so Ising-type behavior is indicated. Field-theoretic approaches support
this picture~\cite{moreira94}.

However, the theoretical arguments are by no means rigorous and have not, so
far, been tested by precise calculations for appropriate models. To do that is
the aim of the researches reported here. We have studied a finely-discretized
version~\cite{azp} of the simplest continuum model (considered by Debye and
H\"uckel in 1923~\cite{fisher-review,weing}, three years before Ising's work),
namely, the \emph{restricted primitive model} (RPM), consisting of $N=N_+ +
N_-$ equisized hard spheres of diameter~$a$, precisely half carrying a
charge~$+q_0$ and half $-q_0$, in a medium (representing a solvent) of
dielectric constant~$D$. At a separation $r \geq a$, like (unlike) ions
interact through the potential $\pm q_0^2/Dr$; thus appropriate reduced
density, $\rho=N/V$ for volume~$V$, and temperature variables are
\begin{equation}
  \rho^* = \rho a^3 \;, \quad
  T^* = k_B T D a / q_0^2 \;, \quad 
  t= (T-T_c)/T_c \;.
  \label{eq:criticalvalues}
\end{equation}

Except at low densities and high temperatures, when the inverse Debye length
$\kappa_D a = (4\pi\rho^*/T^*)^{1/2}$ is small, the RPM is intractable
analytically or via series expansions~\cite{fisher-review,stell94,bekiranov99}.
However, it has been much studied by Monte Carlo (MC)
simulations~\cite{fisher-review,caillol,orkoulas,jcp} which have recently
approached the consensus $T_c^* \simeq 0.049$, $\rho_c^* = 0.060$--$0.085$.
However, these values have been derived by assuming Ising-type criticality: on
that basis Bruce--Wilding extrapolation procedures have been
employed~\cite{caillol,orkoulas} (which, even then, neglect potentially
important, asymmetric `pressure-mixing' terms~\cite{fisher-orkoulas00}.)  It
must be stressed that implementing appropriate finite-size extrapolation
methods constitutes the heart of the computational task since a grand-canonical
(GC) system confined in a simulation `box' of dimensions $L \times L \times L$
(with, say, periodic boundary conditions~\cite{ewald}) \emph{cannot} exhibit a
sharp critical point; a finite canonical system may become critical but can
display \emph{only classical} or vdW behavior~\cite{ofp}.

Thus, while previous RPM simulations~\cite{caillol,orkoulas} demonstrate
\emph{consistency} with Ising (or $n=1$) behavior, \emph{no other} universality
classes are ruled out: see also~\cite{jcp,ofp,camp}. Putative `nearby'
candidates are XY or $n=2$ systems (with $\gamma \simeq 1.316$, $\nu \simeq
0.670$), self-avoiding walks (SAWs, $n=0$: with $\gamma \simeq 1.159$, $\nu
\simeq 0.588$)~\cite{ofp,guida-zj} and long-range, $1/r^{d+\sigma}$ scalar
systems (with $d=3$, $\sigma < 2-\eta$)~\cite{camp,lrfss}. On the other
hand, in a preparatory GCMC study \textbf{II}~\cite[(b)]{ofp} of the hard-core
square-well (HCSW) fluid---for which Ising criticality has long been
anticipated---new, \emph{unbiased}, finite-size extrapolation techniques
enabled the $n=2$ and~$0$ classes to be convincingly excluded.

\begin{figure}
\includegraphics[width=\figurewidth]{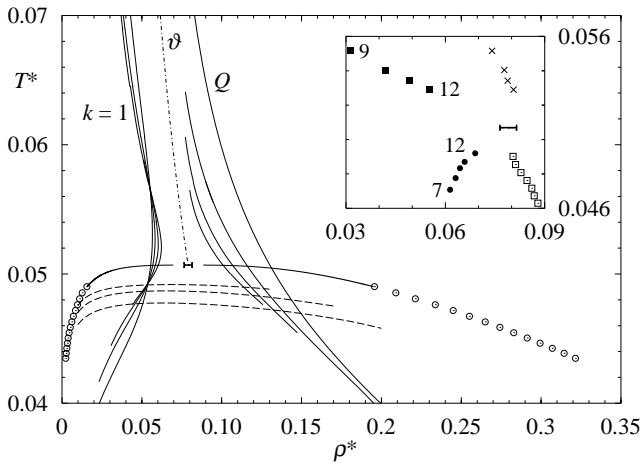}
\caption{Approximate coexistence curve of the RPM in the $(T,\rho)$ plane: open
circles and fitted line. The estimated critical point is shown as an
uncertainty bar.  The dashed curves are loci of $C_V(T)$ maxima at fixed~$\rho$
for $L^* \equiv L/a = 8$, $10$, and~$12$. The loci labeled $k=1$, $\vartheta$,
and~$Q$ are explained in the text. The inset shows the canonical critical
points $T_c^0(L)$, $\rho_c^0(L)$ (squares), and corresponding GC mean densities
$\rho_c^\dagger(L)$ (crosses) for $L^*=9$--$12$, the $C_V(L)$ extrema
$T_c^-(L), \rho_c^-(L)$, for $L^*=7$--$10$ and~$12$ (solid circles), and the
$\sqrt{\rho}$ diameter, $\bar\rho^*_{1/2}(T)$, defined in the text (open
squares).}
\label{fig:coex}
\end{figure}

\paragraph{Present approach}%
We have now applied the methods of \textbf{II} to the RPM; however, the extreme
\emph{asymmetry} of the critical region in the model (see Fig.~\ref{fig:coex})
has demanded further developments. By extending finite-size scaling
theory~\cite{kim} and previous applications of the Binder parameter or
fourth-moment ratio~\cite{lrfss,kim,binder}
\begin{equation}
  Q_L(T; \rho) \equiv \langle m^2\rangle^2/\langle m^4\rangle
  \quad\text{with}\quad m = \rho - \langle\rho\rangle \;,
  \label{eq:Q}
\end{equation}
\cite{note-expectation} to systems lacking symmetry, we have assembled
evidence, outlined below, that excludes not only classical criticality in the
RPM but also the XY and SAW universality classes and $(d=3)$ long-range
Ising criticality with $\sigma \lesssim 1.9$.

Our work employs multihistogram reweighting~\cite{histogram} and a ($\zeta \! =
\! 5$)-level fine-discretization formulation (with a fine-lattice spacing
$a/\zeta$~\cite{azp}). Since $\zeta < \infty$, \emph{non}universal parameters,
such as $T_c^*$, will deviate slightly from their continuum limit
($\zeta\to\infty$)~\cite{azp,note-discretization}; but, at this level, there
are no serious grounds for contemplating changes in universality class. For the
critical parameters we find $T_c^* = 0.05069(2)$ and $\rho_c^*=0.0790(25)$: the
confidence limits in parentheses refer, here and below, to the last decimal
place quoted. The inset in Fig.~\ref{fig:coex} shows how these values are
approached (i) by the canonical values $T_c^0(L)$, $\rho_c^0(L)$ and
$\rho_c^\dagger(L)$ ($=\langle \rho\rangle_{T_c^0(L),\,
\mu_c^0(L)}$~\cite{note-expectation}) derived from the isothermal density
histograms [see \textbf{II}(2.18)--(2.23), Figs. 1,~3], (ii) by $T_c^-(L)$ and
$\rho_c^-(L)$, from the isochoric maxima of $C_V(T; \rho; L)$ [see
Fig.~\ref{fig:coex} and \textbf{II} Sec.~III, Fig.~7], and (iii) by the
$\sqrt{\rho}$ diameter, $\bar\rho_{1/2}(T)$, defined below.

\begin{figure}
\includegraphics[width=\figurewidth]{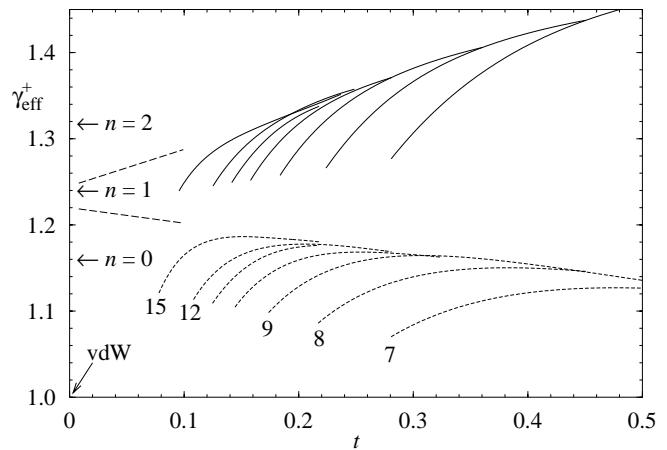}
\caption{Effective susceptibility exponent $\gamma_{\text{eff}}^+(T)$ for
$\rho=\rho_c$ (solid curves) and $\tilde\gamma^+_{\text{eff}}(T)$ on the theta
locus (dashed; see text), for sizes $L^*=7$--$12$ and~$15$. Values for
vdW and for $n=0$, $1$, and~$2$ are marked on the $\gamma$ axis.}
\label{fig:gamma}
\end{figure}

\paragraph{Exponents $\bm\gamma$ and $\bm\nu$}%
Before justifying the precision of our $(T_c, \rho_c)$ estimates, we consider
their implications. The solid curves in Fig.~\ref{fig:gamma} portray the
effective susceptibility exponent $\gamma_{\text{eff}}^+(T; L)$ on the critical
isochore above $T_c$, as derived from $\chi_{NN} \equiv V\langle m^2\rangle =
k_B T \rho^2 K_T$: see \textbf{II}(3.7). Within statistical precision the data
are independent of the $(T_c, \rho_c)$ uncertainties.

Also presented in Fig.~\ref{fig:gamma} are the modified estimators
$\tilde\gamma^+_{\text{eff}}(T)$ [defined as in \textbf{II}(3.7) but with $t$
replacing $t'$] evaluated on the `theta locus,' $\rho_\vartheta(T) =
\rho_c[\vartheta + (1-\vartheta)(T_c/T)]$. This relation approximates an
\emph{effective symmetry locus}~(\textbf{II}) above $T_c$, derived from the
behavior of the isothermal inflection loci $\rho_k(T; L)$, on which $\chi^{(k)}
\equiv \chi_{NN}(T, \rho; L)/\rho^k$ is maximal [see
\textbf{II}(2.26)--(2.32)]. The $k=1$ loci are shown in Fig.~1 for $L^* \equiv
L/a = 6,8,10,12$; the selected value $\vartheta \!= \! 0.20$ corresponds
roughly to $k\simeq 0.60$ (which may be identified with an optimal value: see
\textbf{II} and~\cite{kim}).  However, the variation of the $k$ loci when $L$
increases is significantly more complicated in the RPM than in the
HCSW fluid~[11(c), 23].%\cite[(c)]{jcp} \cite{note-estimates}.
% There is no way to get this cited properly in RevTeX

Extrapolation of the effective susceptibility exponents in Fig.~\ref{fig:gamma}
and those on the $k=0$ locus, etc.~\cite[(c)]{jcp}, to $t=0$ indicates
$\gamma=1.24(3)$, upholding Ising-type behavior while both XY and SAW values
are implausible.

\begin{figure}
\includegraphics[width=\figurewidth]{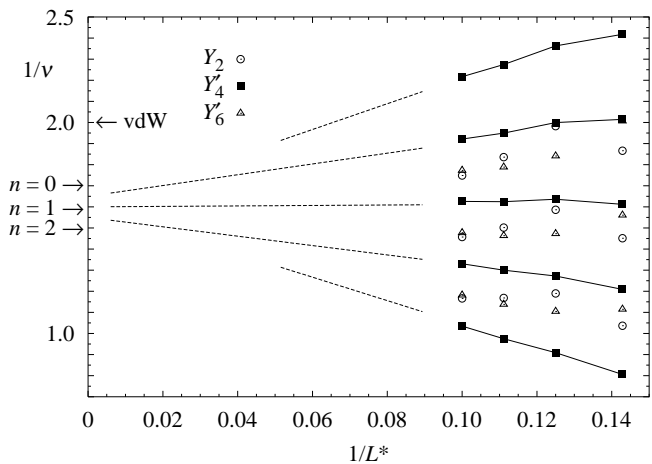}
\caption{Estimation of the correlation exponent~$\nu$ from the deviations
$T_j(L)-T_c$ for various properties $Y_j(T)$ on the critical isochore: see text
and \protect\cite[(c)]{jcp}. Values for $n=0,1,2$, i.e., SAW, Ising, and XY,
and classical (vdW) criticality are indicated.}
\label{fig:nu}
\end{figure}

To determine the exponent~$\nu$ we have examined the peak positions, $T_j(L)$,
of various properties, $Y_j(T; L)$, on the critical isochore.  Finite-size
scaling theory~\cite{kim} yields $\Delta T_j(L) \equiv T_j(L)-T_c \sim
L^{-1/\nu}$: Figure~\ref{fig:nu} demonstrates the estimation of $1/\nu$
(unbiased except for the imposed $T_c$ estimate) from the ratios $\Delta
T_j(L_1) / \Delta T_j(L_2)$ for various $j$ (see \cite[(c)]{jcp}), using an
established approach [see \textbf{I}(7)--(13), Fig.~1; \textbf{II}(3.1)]. The
data indicate $\nu = 0.63(3)$, excluding classical but supportive of Ising
($n=1$) criticality, while $n=2$ and~$0$ seem less probable.

\paragraph{Estimation of $\bm{T_c^*}$}%
Consider, now, $Q_L(T; \rho)$ in~(\ref{eq:Q}), when $L\to\infty$. In \emph{any
single}-phase region of the $(T, \rho)$ plane $Q_L \to\frac{1}{3}$, indicative
of Gaussian fluctuations about $\langle \rho\rangle$; conversely, \emph{within}
a two-phase region, $\rho_-(T) < \rho < \rho_+(T)$, one finds $Q_L\to 1$
\emph{on} the \emph{diameter}, $\bar\rho(T) \equiv \frac{1}{2}(\rho_- +
\rho_+)$ for $T<T_c$, while, more generally,
\begin{equation}
1 \geq Q_\infty(T; \rho) = 1 - 4y^2/(1+6y^2+y^4) > \textstyle{\frac{1}{2}} \;,
\label{eq:qcoex}
\end{equation}
where $|y|=2|\rho-\bar\rho(T)|/(\rho_- + \rho_+) < 1$. Finally, \emph{at
criticality}, $Q_L(T_c; \rho_c)$ approaches a \emph{universal value} $Q_c$
which, for cubic boxes with periodic boundary conditions, is $Q_c =
0.4569\cdots$ for classical (vdW)~\cite[(b)]{binder} or $\infty$-range
systems~\cite[(c)]{binder} but $Q_c(n \! =\! 1) = 0.6236(2)$ for
Ising~\cite[(d),(e)]{binder} and $Q_c(n \! = \! 2) = 0.8045(1)$ for
XY~\cite[(f)]{binder} systems, while $Q_c(n \! =\!  0) = 0$~\cite[(b)]{binder}.
For long-range, $1/r^{3+\sigma}$ systems, $Q_c(\sigma)$ and also
$\gamma(\sigma)$, increase almost linearly from vdW to Ising values in the
interval $\frac{3}{2} \leq \sigma \leq (\gamma/\nu)_{n=1} \simeq 1.96_6$ with
$Q_c(\sigma \! = \!  1.9) \simeq 0.600$ and $\gamma(\sigma \! = \! 1.9) \simeq
1.20_5$~\cite[(b)]{lrfss}.

The result~(\ref{eq:qcoex}) leads us to propose $Q$-loci, $\rho_Q(T; L)$, on
which $Q_L(T; \rho)$ is maximal at fixed~$T$. For $T<T_c$ these loci are
observed to approach the diameter $\bar\rho(T)$ when $L$ increases.  (For $T
\lesssim T_c$, but \emph{not} above $T_c$, the $Q$-loci also follow the $k=0$
loci quite closely.)

\begin{figure}
\includegraphics[width=\figurewidth]{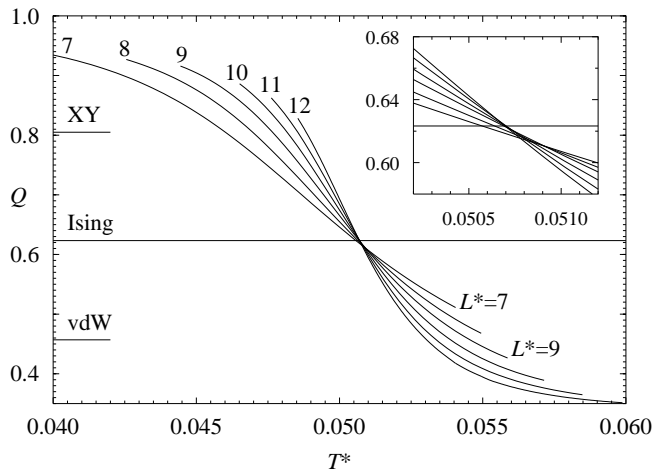}
\caption{Plots of $Q_L(T; \rho)$ on the $Q$-loci, $\rho_Q(T; L)$, providing
estimates for $T_c$ and $Q_c$. Classical, XY and Ising values of $Q$ are
shown.}
\label{fig:Q}
\end{figure}

Figure~\ref{fig:Q} displays $Q_L(T; \rho)$ on the $Q$-loci $\rho_Q(T; L)$, for
$L^*=7$--$12$. As often seen in plots for \emph{symmetric}
systems~\cite{binder}, inflection points and successive intersections,
$T_Q(L)$, almost coincide! Scaling yields $Q_L(T_c; \rho_c) \sim
L^{-\theta/\nu}$ and $|T_c-T_Q(L)| \sim L^{-\varphi}$ with $\varphi =
(1+\theta)/\nu$, where $\theta$ $(=\omega\nu)$ is the leading
correction-to-scaling exponent; for classical and Ising criticality one has
$(\theta/\nu, \varphi) = (1, 3)$, $\simeq (0.82, 2.41)$~\cite{guida-zj}. With
this guidance, the large-scale inset in Fig.~\ref{fig:Q} leads to our estimate
$T_c^* \simeq 0.05069(2)$ but also yields $Q_c \simeq 0.624(2)$: this is
surprisingly close to the Ising value~\cite{note-q12} and far from the vdW, XY,
and SAW values---an unexpected bonus! Likewise, $1/r^{3+\sigma}$ effective
potentials with $\sigma \leq 1.9$ are excluded.

\begin{figure}
\includegraphics[width=\figurewidth]{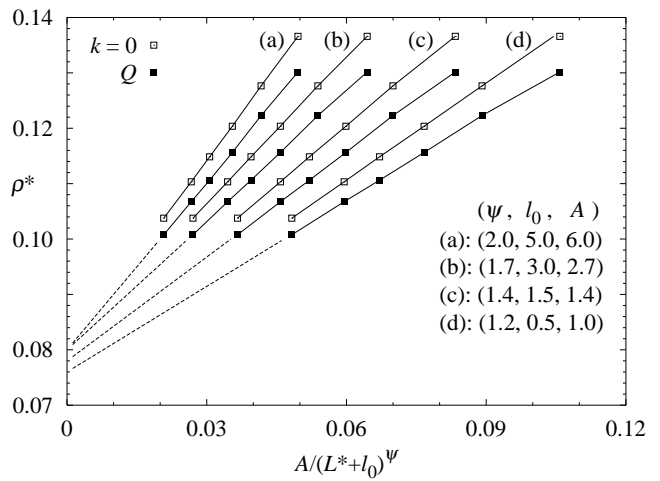}
\caption{Estimation of $\rho_c^*$ from plots of $(k=0)$ and $Q$ locus values at
$T_c$ (open and solid squares) vs $A/(L^*+\ell_0)^\psi$ for various values of
$\psi$ and optimal shifts~$\ell_0$. The scale parameter~$A$ has been invoked
merely for graphical clarity.  Note $\psi < 1.6$ requires smaller shifts
tending to exclude vdW criticality ($\psi=2$).}
\label{fig:rhoc}
\end{figure}

\paragraph{Estimation of $\bm{\rho_c}$}%
Finally, we examine $\rho_0^c(L)$ and $\rho_Q^c(L)$, i.e., the ($k=0$) and $Q$
loci intersections with the estimated critical isotherm, $T=T_c$.  According to
scaling, the deviations, $\Delta \rho_0^c$ and $\Delta \rho_Q^c$, decay as
$L^{-\psi}$ with $\psi=(1-\alpha)/\nu$~\cite{kim}, so we may suppose $1.2 <
\psi \leq 2$~\cite{guida-zj}.  Figure~\ref{fig:rhoc} displays the deviations vs
$L^{-\psi}$ for $\psi = 1.2$, $1.4$, $1.7$ and~$2$ with `$\ell_0$ shifts'
[\textbf{I}(19), Fig.~2; \textbf{II}(3.1)] chosen to provide linear plots.
From these and further plots~\cite[(c)]{jcp} we conclude $\rho_c^*=0.0790(25)$.

In further support of our $\rho_c$ estimate, we mention first that when the
coexistence curve, $\rho_\pm(T)$, is plotted vs $\sqrt{\rho^*}$---as is
reasonable since all powers $\rho^{j/2}$ for integral~$j$ appear in virial
expansions for the RPM~\cite{bekiranov99}---it becomes markedly more
symmetrical [resembling $(\rho, T)$ plots for the HCSW and other simple
fluids]. Then, the \emph{corresponding} diameter,
$\sqrtsign[\bar\rho^*_{1/2}(T)] = \frac{1}{2}(\sqrt{\rho^*_-} +
\sqrt{\rho^*_+})$, is only mildly curved and naive extrapolations to $T_c$
yield $\rho_c^* = 0.078(4)$.

\paragraph{In conclusion}%
By implementing recently tested~\cite{ofp} and newly devised extrapolation
techniques for nonsymmetric critical systems, our extensive grand-canonical
Monte Carlo simulations for the RPM have provided, \emph{in toto}, convincing
evidence to exclude classical, XY ($n=2$), or SAW ($n=0$) critical behavior as
well as long-range (effective) Ising interactions decaying more slowly than
$1/r^{4.90}$.  Rather, the estimates for the exponents $\nu$ and $\gamma$, and
for the critical fourth-moment ratio, $Q_c$, point to standard, short-range
Ising-type criticality.  Studies underway~\cite[(c)]{jcp} should provide
further confirmation and additional quantitative results, such as the scale,
$R_0$, of the equivalent single-component short-range attractions generated by
the RPM near criticality.

\begin{acknowledgments}
  We are indebted to Young C. Kim for extensive assistance in the numerical
  analysis, for his elucidation of the finite-size scaling properties of
  asymmetric fluid criticality~\cite{kim}, and his discovery of effective
  estimators for $\nu$.  The support of the National Science Foundation
  (through Grant No.\ 99-81772, MEF) and the Department of Energy, Office of
  Basic Energy Sciences (DE-FG02-01ER15121, AZP) is gratefully acknowledged.
\end{acknowledgments}


\begin{thebibliography}{99}
  
\bibitem{fisher-review} See (a) M. E. Fisher, J. Stat. Phys. \textbf{75}, 1
  (1994), (b) J. Phys. Condens. Matter \textbf{8}, 9103 (1996), and references
  therein.
  
\bibitem{weing} H. Weing\"artner and W. Schr\"oer, Adv. Chem. Phys.
  \textbf{116}, 1 (2001).

\bibitem{stell94} See G. Stell, J. Stat. Phys. \textbf{78}, 197 (1994).
  
\bibitem{fisher-lee96} M. E. Fisher and B. P. Lee, Phys. Rev. Lett.
  \textbf{77}, 3561 (1996).
  
\bibitem{fisher-levin93} M. E. Fisher and Y. Levin, Phys. Rev. Lett.
  \textbf{71}, 3826 (1993).
  
\bibitem{moreira94} See, e.g., A. G. Moreira, M. M. Telo da Gama and M. E.
  Fisher, J. Chem. Phys. \textbf{110}, 10058 (1999) and references therein.

\bibitem{azp} (a) A. Z. Panagiotopoulos and S. K. Kumar,
  Phys. Rev. Lett. \textbf{83}, 2981 (1999); (b) A. Z. Panagiotopoulos,
  J. Chem. Phys. \textbf{112}, 7132 (2000); (c) A. Z. Panagiotopoulos,
  J. Chem. Phys. [in press].
  
\bibitem{bekiranov99} S. Bekiranov and M. E. Fisher, Phys. Rev. E \textbf{59},
  492 (1999).

\bibitem{caillol} J. M. Caillol, D. Levesque and J. J. Weis,
  Phys. Rev. Lett. \textbf{77}, 4039 (1996); J. Chem. Phys. \textbf{107}, 1565
  (1997).

\bibitem{orkoulas} G. Orkoulas and A. Z. Panagiotopoulos,
  J. Chem. Phys. \textbf{101}, 1452 (1994); \textbf{110}, 1581 (1999);
  Q. Yan and J. J. de Pablo, J. Chem. Phys. \textbf{111}, 9509 (1999).
  
\bibitem{jcp} E. Luijten, M. E. Fisher and A. Z. Panagiotopoulos, (a) J. Chem.
  Phys. \textbf{114}, 5468 (2001); (b) a preliminary account of the present
  work was presented at \textsc{STATPHYS 21} in Cancun, Mexico, on 16 July
  2001; (c) to be published.

\bibitem{fisher-orkoulas00} M. E. Fisher and G. Orkoulas,
  Phys. Rev. Lett. \textbf{85}, 696 (2000).
  
\bibitem{ewald} We use Ewald summations with conducting boundary conditions
  ($\varepsilon_\infty \to \infty$)~\protect\cite[(c)]{azp}, a screening
  parameter $\kappa = 6/L$, chosen so as to ensure a sufficiently fast decay of
  the real-space charge distribution, and 1152 wave vectors (to accomodate the
  relatively rapid variation of the charge distribution).
  
\bibitem{ofp} (a) G. Orkoulas, A. Z. Panagiotopoulos and M. E. Fisher, Phys.
  Rev. E \textbf{61}, 5930 (2000); (b) G. Orkoulas, M. E. Fisher and A. Z.
  Panagiotopoulos, Phys. Rev. E \textbf{63}, 051507 (2001); to be denoted
  \textbf{I} and \textbf{II}.

\bibitem{camp} P. J. Camp and G. N. Patey, J. Chem. Phys. \textbf{114}, 399
  (2001).

\bibitem{guida-zj} R. Guida and J. Zinn-Justin, J. Phys. A \textbf{31}, 8103
  (1998).
  
\bibitem{lrfss} E. Luijten, (a) Phys. Rev. E \textbf{60}, 7558 (1999); (b)
  \textit{Interaction range, Universality and the Upper Critical Dimension}
  (Delft Univ. Press, Delft, 1997).

\bibitem{kim} Y. C. Kim and M. E. Fisher (to be published).
  
\bibitem{binder} (a) K. Binder, Z. Phys. B \textbf{43}, 119 (1981); (b) E.
  Br\'ezin and J. Zinn-Justin, Nucl. Phys. B \textbf{257}, 867 (1985); (c) E.
  Luijten and H. W. J. Bl\"ote, Int. J. Mod. Phys. C \textbf{6}, 359 (1995);
  (d) H. W. J. Bl\"ote, E. Luijten and J. R. Heringa, J. Phys. A \textbf{28},
  6289 (1995); (e) H. W. J. Bl\"ote, L. N. Shchur and A. L. Talapov, Int. J.
  Mod. Phys. C \textbf{10}, 1137 (1999); (f) M. Campostrini \emph{et al.},
  Phys. Rev.  B \textbf{63}, 214503 (2001).

\bibitem{note-expectation} All expectation values, $\langle \bm{\cdot}
  \rangle$, pertain to finite GC systems at fixed~$T$ and chemical
  potential~$\mu$, chosen, when appropriate, to provide a mean density $\langle
  \rho\rangle$ which we denote merely by $\rho$ when used as a variable, as in
  $Q_L(T; \rho)$.

\bibitem{histogram} A. M. Ferrenberg and R. H. Swendsen,
  Phys. Rev. Lett. \textbf{63}, 1195 (1989).

\bibitem{note-discretization} Compare J. M. Romero-Enrique \emph{et al.}, Phys.
  Rev. Lett. \textbf{85}, 4558 (2000), using $\zeta=10$, with Q. L. Yan and J.
  J. de Pablo, Phys. Rev. Lett. \textbf{86}, 2054 (2001) and see
  \protect\cite[(c)]{azp}.

\bibitem{note-estimates} For this reason the estimates quoted in the \emph{Note
  added in proof} in \protect\cite[(a)]{jcp} are inaccurate.
  
\bibitem{note-q12} A similar analysis of $Q_L^{(1,2)} \equiv \langle
  |m|\rangle^2/\langle m^2\rangle$ on the RPM $Q$-loci confirms $T_c^*$ and
  yields $Q_c^{(1,2)} \simeq 0.807$ close to the corresponding Ising value
  $0.8070(9)$ (E. Luij\-ten, unpublished).

\end{thebibliography}
\end{document}